\documentclass[english]{revtex4}
\usepackage[T1]{fontenc}
\usepackage[latin9]{inputenc}
\usepackage{color}
\usepackage{amssymb}

\makeatletter

\providecommand{\boldsymbol}[1]{\mbox{\boldmath $#1$}}

\usepackage{babel}
\makeatother

\newcommand{\be}{\begin{equation}}
\newcommand{\ee}{\end{equation}}
\newcommand{\bea}{\begin{eqnarray}}
\newcommand{\eea}{\end{eqnarray}}

\newcommand{\gapp}{\mathrel{\raise.3ex\hbox{$>$}\mkern-14mu
              \lower0.6ex\hbox{$\sim$}}}
\newcommand{\lapp}{\mathrel{\raise.3ex\hbox{$<$}\mkern-14mu
              \lower0.6ex\hbox{$\sim$}}}

\newcommand\lsim{\lesssim}
\newcommand\gsim{\gtrsim}

\renewcommand\[{\left[}
\renewcommand\]{\right]}

\newcommand\eqreff[1]{(\ref{#1})}

\def\calp{{\cal P}}

\def\calpz{{\calp_\zeta}}

\newcommand\bfd{{\mathbf d}}

\newcommand\bfk{{\mathbf k}}

\newcommand\sub[1]{_{\rm #1}}
\newcommand\su[1]{^{\rm #1}}

\newcommand\mtwo{^{-2}}

\newcommand\half{^{1/2}}

\newcommand{\fnl}{f\sub{NL}}

\begin{document}

\title{Anisotropic non-Gaussianity from vector field perturbations}

\author{Mindaugas Kar\v{c}iauskas}

\email{m.karciauskas@lancaster.ac.uk}

\affiliation{Department of Physics, Lancaster University, Lancaster LA1 4YB, UK}

\author{Konstantinos Dimopoulos}

\email{k.dimopoulos1@lancaster.ac.uk}

\affiliation{Department of Physics, Lancaster University, Lancaster LA1 4YB, UK}

\author{David H. Lyth}

\email{d.lyth@lancaster.ac.uk}

\affiliation{Department of Physics, Lancaster University, Lancaster LA1 4YB, UK}

\begin{abstract}
We suppose that a vector field perturbation causes part of the 
primordial curvature perturbation. The non-Gaussianity parameter $\fnl$
is then, in general, statistically anisotropic. We calculate its form and 
magnitude in the curvaton scenario and in the end-of-inflation scenario. 
We show that this anisotropy could easily  be observable.
\end{abstract}
\maketitle

\section{Introduction}

The primordial curvature perturbation $\zeta$ provides one of the
few windows available to the early Universe, and is the subject of
intense interest at present. It may be that $\zeta$ is practically
Gaussian. Then its Fourier components will have practically no correlation,
except for the one required by the reality condition. The latter is
defined by the two-point correlator, specified by the spectrum $\calpz(k)$
where $k$ is the wavenumber. On cosmological scales, the observed
CMB anisotropy gives an almost scale-independent value $\calpz\simeq(5\times10^{-5})^{2}$.

According to typical scenarios for the generation of $\zeta$, the
principle signal for non-Gaussianity would be the three point correlator,
specified by the non-linearity parameter $\fnl$. At present there
is only an upper bound $|\fnl|\lsim100$. Over the next few years,
the bound will go down to $|\fnl|\lsim5$ or so if there is no detection.

During inflation, the vacuum fluctuation of each canonically normalized
light scalar field becomes a classical perturbation, with a nearly
scale-independent and Gaussian spectrum. It is usually supposed that
one or more of these perturbations is responsible for the primordial
curvature perturbation $\zeta$. In this case, $\zeta$ is statistically
homogeneous and isotropic, which means that its correlators are invariable
under translations and rotations. Then $\calpz(k)$ depends only on
the magnitude of a momentum (wave vector) $\bfk$, and $\fnl(k_{1},k_{2},k_{3})$
depends on the lengths of the sides of a triangle.

Most of the proposals for generating $\zeta$ from the scalar field
perturbations belong to one of two broad classes, which are distinguished
by their prediction for $\fnl$. If $\zeta$ is generated during single-field
inflation one generally has $|\fnl|\lsim10\mtwo$. More strongly,
this bound applies to all single-field models in the squeezed configuration
$k_{1}\simeq k_{2}\gg k_{3}$. If instead $\zeta$ is generated at
or after the end of inflation, $\fnl$ becomes almost scale-independent.
In this case the prediction for $|\fnl|$ is usually at least of order
1 and can be as big or bigger than the observational bound.

Under the assumption of statistical isotropy, the constraints on $\fnl$
from current observation at the {\em two}-$\sigma$ level are \cite{wmap}
\be -9 < \fnl\su{local} < 111,\qquad{}-151<\fnl\su{equil}
< 253 , \ee where the label `local' can be taken to mean the squeezed
configuration $k_{1}\simeq k_{2}\ll k_{3}$ and the label `equil'
can be taken to mean the equilateral configuration $k_{1}=k_{2}=k_{3}$.
The first result might be regarded as weak evidence for a $\fnl\su{local}\gg1$,
which if confirmed would rule out the generation of $\zeta$ during
single-field inflation.

Recently, it has been suggested that the perturbation of some {\em
vector} field may generate part or even all of the curvature perturbation.
Such a thing is possible because the vacuum fluctuation of a vector
field can generate an almost scale-independent and Gaussian spectrum
although the condition for that to happen is rather special (e.g. it doesn't
happen for a canonically normalized light vector field.) The contribution
to $\zeta$ from a vector field perturbation is statistically homogeneous
but {\em not} in general statistically isotropic. Then the spectrum
may depend on the direction of $\bfk$, and the bispectrum may depend
on the orientation of the triangle of vectors $(k_{1},k_{2},k_{3})$.

Observational bounds on to statistical anisotropy have not received
much attention, and are not mentioned in the otherwise comprehensive
analysis by the WMAP team \cite{wmap}. As far as we know, the only
available result \citep{ge} concerns the spectrum. It is parametrized
in the form \be \calpz(\bfk) = \[
1+g(\hat{\bfd}\cdot\hat{\bfk})^{2}\]
 \calpz\su{iso}(k) , \label{specanis} \ee where \mbox{$k=|\bfk|$} and
the hats denote
unit vectors. After taking account of all possible uncertainties,
the conclusion from this study is $g\lsim0.31$ or so. We will adopt
a bound $g\lsim0.1$ for definiteness.

Using the $\delta N$ formalism \citep{Starobiskii(1985),Sasaki_Stewart(1996),Lyth_Rodriguez(2005),Lyth_etal(2005)}, one can write general formulas 
for the contribution of a vector field perturbation to the spectrum
and the bispectrum. 
They were evaluated in Ref.~\cite{Yokoyama_Soda(2008)}
for the case that $\zeta$ is generated at the end of inflation, and
in Ref.~\citep{Dimopoulos_etal(2008)} for the case that $\zeta$
is generated by the curvaton mechanism. (Vector field inflation was
also considered in Ref.~\citep{Dimopoulos_etal(2008)} but we shall
not consider it here.) It was found that the contribution to $\calpz$
spectrum is of the form shown in Eq.~\eqreff{specanis}. An analogous form for
the contribution to $\fnl$ is to obtained in this work.
These are the tree-level contributions. The 1-loop contribution to
$\calpz$ is given in Ref.~\citep{Dimopoulos_etal(2008)}
but the 1-loop contribution to $\fnl$ is not known
at the time of writing. 

{}From the formula for $\calpz$, one sees that its statistical anisotropy
could easily be as big or bigger than the observational bound. The
purpose of this paper is to consider also $\fnl$. We want to know
if a vector field contribution to $\fnl$ could be big enough to observe,
bearing in mind that its contribution to $\calpz$ should respect
the observational bound. We answer this question in the affirmative,
and give explicit formulas for the dependence of $\fnl$ on the orientation
of the triangle.

\section{\label{sec:general_fomulae}$\boldsymbol{f_{\mathrm{NL}}}$ including
vector perturbations}

The evolution of the curvature perturbation $\zeta$ on superhorizon
scales is most readily described using a separate universe approach
\citep{Starobiskii(1985),Sasaki_Stewart(1996),Lyth_Rodriguez(2005),Lyth_etal(2005)}.
In a recent paper \citep{Dimopoulos_etal(2008)} the formalism was
extended to take into account the possible statistical anisotropy
in $\zeta$, where it was shown that once one includes perturbations
of the vector field, the resulting curvature perturbation can be calculated
up to quadratic terms using the following equation\begin{equation}
\zeta\left(\mathbf{x}\right)=N_{\phi}\delta\phi+N_{i}^{A}\delta A_{i}+\frac{1}{2}N_{\phi\phi}\left(\delta\phi\right)^{2}+\frac{1}{2}N_{\phi i}^{A}\delta\phi\delta A_{i}+\frac{1}{2}N_{ij}^{A}\delta A_{i}\delta A_{j},\label{eq:zeta_x}\end{equation}
where $N$ is the number of e-folds of expansion of the unperturbed
universe, the lower case roman letters denote spatial indices and
Einstein summation over those indices is assumed. The derivatives
of $N$ with respect to the fields are denoted as 
\begin{equation}
N_{\phi}\equiv\frac{\partial N}{\partial\phi},\;\; 
N_{i}^{A}\equiv\frac{\partial N}{\partial A_{i}},\;\; 
N_{\phi\phi}\equiv\frac{\partial^{2}N}{\partial\phi^{2}},\;\; 
N_{\phi i}^{A}\equiv\frac{\partial^{2}N}{\partial\phi\partial A_{i}},
\;\;\mathrm{and}\;\; 
N_{ij}^{A}\equiv\frac{\partial^{2}N}{\partial A_{i}\partial A_{j}}.
\end{equation}
 To obtain Eq.(\ref{eq:zeta_x}) it was assumed that the anisotropy
of the expansion of the Universe is negligible. Even for the vector
fields having a considerable contribution to the curvature perturbation
the isotropic expansion of the Universe can be achieved at least in
several ways, for example with the oscillating massive vector field
\citep{Dimopoulos2006,Dimopoulos2007,Dimopoulos_Karciauskas(2008)},
using a triad of orthogonal vectors \citep{Bento_etal(1993)}, a large
number of identical randomly oriented vector fields \citep{Mukhanov_etal(2008)}
or if the contribution of the vector field(s) to the total energy
density is negligible \citep{Yokoyama_Soda(2008)}. However, in general
perturbations generated by vector fields will induce statistical anisotropy
in $\zeta$. In Eq.(\ref{eq:zeta_x}) we assume only a single scalar
field $\phi$ and a single vector field~$A_{\mu}$.

We define the power spectrum and the bispectrum through the Fourier
modes of $\zeta$ as:\begin{equation}
\left\langle \zeta\left(\mathbf{k}_{1}\right),\zeta\left(\mathbf{k}_{2}\right)\right\rangle \equiv\left(2\pi\right)^{3}\delta^{(3)}\left(\mathbf{k}_{1}+\mathbf{k}_{2}\right)\frac{2\pi^{2}}{k^{3}}\mathcal{P}_{\zeta}\left(\mathbf{k}_{1}\right),\end{equation}
 \begin{equation}
\left\langle \zeta\left(\mathbf{k}_{1}\right)\zeta\left(\mathbf{k}_{2}\right)\zeta\left(\mathbf{k}_{3}\right)\right\rangle \equiv\left(2\pi\right)^{3}\delta^{\left(3\right)}\left(\mathbf{k}_{1}+\mathbf{k}_{2}+\mathbf{k}_{3}\right)B_{\zeta}\left(\mathbf{k}_{1},\mathbf{k}_{2},\mathbf{k}_{3}\right),\end{equation}
 where normalization of Fourier components is chosen to be such that\begin{equation}
\zeta\left(\mathbf{k}\right)\equiv\int\zeta\left(\mathbf{x}\right)\mathrm{e}^{-i\mathbf{k}\cdot\mathbf{x}}\mathrm{d}\mathbf{x}.\end{equation}
 Note that the power spectrum and the bispectrum are dependent on
the direction of $\mathbf{k}$. The bispectrum $B_{\zeta}\left(\mathbf{k}_{1},\mathbf{k}_{2},\mathbf{k}_{3}\right)$
can be further separated into three parts: one due to perturbations
in the scalar field, another part due to the vector field and a mixed
term:\begin{eqnarray}
B_{\phi}\left(\mathbf{k}_{1},\mathbf{k}_{2},\mathbf{k}_{3}\right) & \equiv & N_{\phi}^{2}N_{\phi\phi}\left[\frac{4\pi^{4}}{k_{1}^{3}k_{2}^{3}}\mathcal{P}_{\phi}\left(k_{1}\right)\mathcal{P}_{\phi}\left(k_{2}\right)+\mathrm{c.p.}\right],\nonumber \\
B_{\phi A}\left(\mathbf{k}_{1},\mathbf{k}_{2},\mathbf{k}_{3}\right) & \equiv & -\frac{1}{2}N_{\phi}N_{\phi,i}^{A}\left[\frac{4\pi^{4}}{k_{1}^{3}k_{2}^{3}}\mathcal{P}_{\phi}\left(k_{1}\right)\mathcal{M}_{i}\left(\mathbf{k}_{2}\right)+\mathrm{5\, perm.}\right],\label{eq:bispectra_general}\\
B_{A}\left(\mathbf{k}_{1},\mathbf{k}_{2},\mathbf{k}_{3}\right) & \equiv & \frac{4\pi^{4}}{k_{1}^{3}k_{2}^{3}}\mathcal{M}_{i}\left(\mathbf{k}_{1}\right)N_{ij}^{A}\mathcal{M}_{j}\left(\mathbf{k}_{2}\right)+\mathrm{c.p.},\nonumber \end{eqnarray}
 where {}``c.p'' stands for {}``cyclic permutations'' and $k_{1}$,
$k_{2}$, $k_{3}$ are the moduli of the vectors $\mathbf{k}_{1}$,
$\mathbf{k}_{2}$ and $\mathbf{k}_{3}$.

The power spectrum $\mathcal{P}_{\phi}\left(k\right)$ in the above
equations depends only on the modulus of $\mathbf{k}$ because we
assumed that the expansion during inflation is isotropic. The vector
$\mathcal{M}_{i}\left(\mathbf{k}\right)$ characterizes perturbations
of the vector field:
%
\begin{equation}
\mbox{\boldmath $\mathcal{M}$}\left(\mathbf{k}\right)\equiv
\mathcal{P}_{+}\left(k\right)N_{A}
\left[\hat{\bf N}^{A}+
p(k)\hat{\bf k}\left(\hat{\mathbf{k}}\cdot\hat{\mathbf{N}}^{A}\right)
+iq(k)\,\hat{\mathbf{k}}\times\hat{\mathbf{N}}^{A}\right].
\label{eq:M_def}
\end{equation}
 In this expression $N_{A}$ is the modulus of the vector $\mathbf{N}^{A}$,
$\hat{\mathbf{N}}^{A}$ and $\hat{\mathbf{k}}$ are unit vectors defined
by $\hat{\mathbf{N}}^{A}\equiv\mathbf{N}^{A}/N_{A}$ and $\hat{\mathbf{k}}=\mathbf{k}/k$.
The power spectrum for the longitudinal component is denoted by $\mathcal{P}_{0}\left(k\right)$
while $\mathcal{P}_{+}\left(k\right)$ and $\mathcal{P}_{-}\left(k\right)$
are the parity conserving and violating power spectra defined by\begin{equation}
\mathcal{P}_{\pm}
\equiv\frac{1}{2}
(\mathcal{P}_{R}
\pm\mathcal{P}_{L}
)
,\label{eq:Ppm_def}\end{equation}
with $\mathcal{P}_{R}(k)$ and $\mathcal{P}_{L}(k)$ denoting the power
spectra for the transverse components with right-handed and left-handed
polarizations. Also we have defined $p(k)$ and $q(k)$ as
\begin{equation}
p
\equiv\frac{\mathcal{P}_{0}
-\mathcal{P}_{+}
}{\mathcal{P}_{+}
}\quad\mathrm{and}\quad q
\equiv\frac{\mathcal{P}_{-}
}{\mathcal{P}_{+}
}.\label{eq:g_h_def}\end{equation}
Because of the isotropic expansion during inflation the power
spectra $\mathcal{P}_{0}$ and $\mathcal{P}_{\pm}$
depend only on the modulus of~${\bf k}$.

The curvature perturbation power spectrum $\mathcal{P}_{\zeta}\left(\mathbf{k}\right)$
may be separated into isotropic and anisotropic parts \citep{Ackerman_etal(2007)}:\begin{equation}
\mathcal{P}_{\zeta}\left(\mathbf{k}\right)=
\mathcal{P}_{\zeta}^{\mathrm{iso}}(k)\left[1+g\left(k\right)\left(\hat{\mathbf{N}}^{A}\cdot\hat{\mathbf{k}}\right)^{2}\right],\label{eq:Pzeta_definition}\end{equation}
 where the amount of anisotropy at each scale is parametrized by
\begin{equation}
g
\equiv N_{A}^{2}\;\frac{\mathcal{P}_{0}
-\mathcal{P}_{+}
}{\mathcal{P}_{\zeta}^{\mathrm{iso}}
}.\label{eq:miu_def}\end{equation}
 The isotropic part of the spectrum is\begin{equation}
\mathcal{P}_{\zeta}^{\mathrm{iso}}
\equiv N_{\phi}^{2}\mathcal{P}_{\phi}
+N_{A}^{2}\mathcal{P}_{+}
.\label{eq:Piso_definition}\end{equation}

In this paper we will calculate $f_{\mathrm{NL}}$ for both configurations:
equilateral, in which $k_{1}=k_{2}=k_{3}$, and squeezed, in which
$k_{1}\simeq k_{2}\gg k_{3}$. In the equilateral configuration the bispectra
from Eqs.~(\ref{eq:bispectra_general}) become\begin{eqnarray}
\mathcal{B}_{\phi}^{\mathrm{equil}}\left(\mathbf{k}_{1},\mathbf{k}_{2},\mathbf{k}_{3}\right) & = & 3N_{\phi}^{2}N_{\phi\phi}\mathcal{P}_{\phi}^{2}\left(k_{1}\right),\nonumber \\
\mathcal{B}_{A\phi}^{\mathrm{equil}}\left(\mathbf{k}_{1},\mathbf{k}_{2},\mathbf{k}_{3}\right) & = & -N_{\phi}N_{\phi,i}^{A}\mathcal{P}_{\phi}\left(k_{1}\right)\left[\mathcal{M}_{i}\left(\mathbf{k}_{1}\right)+\mathcal{M}_{i}\left(\mathbf{k}_{2}\right)+\mathcal{M}_{i}\left(\mathbf{k}_{3}\right)\right],\label{eq:BA_def}\\
\mathcal{B}_{A}^{\mathrm{equil}}\left(\mathbf{k}_{1},\mathbf{k}_{2},\mathbf{k}_{3}\right) & = & \mathcal{M}_{i}\left(\mathbf{k}_{1}\right)N_{ij}^{A}\mathcal{M}_{j}\left(\mathbf{k}_{2}\right)+\mathrm{c.p.},\nonumber 
\label{calB}
\end{eqnarray}
where we have defined for the equilateral configuration \begin{equation}
\mathcal{B}_{\zeta}^{\mathrm{equil}}\left(\mathbf{k}_{1},\mathbf{k}_{2},\mathbf{k}_{3}\right)\equiv\left(\frac{k_{1}^{3}}{2\pi^{2}}\right)^{2}B_{\zeta}^{\mathrm{equil}}\left(\mathbf{k}_{1},\mathbf{k}_{2},\mathbf{k}_{3}\right),\label{eq:curlyB_equil_def}\end{equation}
and 
$\mathcal{B}_{\zeta}^{\mathrm{equil}}
=\mathcal{B}_{\phi}^{\mathrm{equil}}
+\mathcal{B}_{A\phi}^{\mathrm{equil}}
+\mathcal{B}_{A}^{\mathrm{equil}}
$.
In this case the non-linearity parameter $f_{\mathrm{NL}}^{\mathrm{equil}}$
is expressed through the power spectrum and the bispectrum as:\begin{equation}
\frac{6}{5}f_{\mathrm{NL}}^{\mathrm{equil}}=\frac{\mathcal{B}_{\zeta}^{\mathrm{equil}}\left(\mathbf{k}_{1},\mathbf{k}_{2},\mathbf{k}_{3}\right)}{3\mathcal{P}_{\zeta}^{\mathrm{iso}}\left(k\right)^{2}}.\label{eq:fNL_definition1}\end{equation}
 Observations give a limit on the anisotropy $g\lesssim0.1$
\citep{ge}. 
 Therefore, since the anisotropic contribution to the curvature perturbation
is subdominant compared to the isotropic one, we have included only
$\mathcal{P}_{\zeta}^{\mathrm{iso}}$ into the above expression of~$f_{\mathrm{NL}}^{\mathrm{equil}}$.

For the squeezed configuration two of the vectors have almost identical lengths but opposite directions, 
$\mathbf{k}_{1}\simeq-\mathbf{k}_{2}$,
but the third vector $\mathbf{k}_{3}$ is of much smaller modulus than the other
two and almost perpendicular to them. For this configuration Eqs.~(\ref{eq:bispectra_general})
take the form
\begin{eqnarray}
\mathcal{B}_{\phi}^{\mathrm{local}}\left(\mathbf{k}_{1},\mathbf{k}_{2},\mathbf{k}_{3}\right) & = & 2N_{\phi}^{2}N_{\phi\phi}\mathcal{P}_{\phi}\left(k_{1}\right)\mathcal{P}_{\phi}\left(k_{3}\right),\nonumber \\
\mathcal{B}_{A\phi}^{\mathrm{local}}\left(\mathbf{k}_{1},\mathbf{k}_{2},\mathbf{k}_{3}\right) & = & -N_{\phi}N_{\phi,i}^{A}\{\mathcal{P}_{\phi}\left(k_{1}\right)\mathcal{M}_{i}\left(\mathbf{k}_{3}\right)+\mathcal{P}_{\phi}\left(k_{3}\right) \mathrm{Re} \left[ \mathcal{M}_{i}\left(\mathbf{k}_{1}\right) \right] \},\label{eq:Bs_local}\\
\mathcal{B}_{A}^{\mathrm{local}}\left(\mathbf{k}_{1},\mathbf{k}_{2},\mathbf{k}_{3}\right) & = & 2\, \mathrm{Re} \left[ \mathcal{M}_{i}\left(\mathbf{k}_{1}\right) \right]\, N_{ij}^{A} \, \mathrm{Re} \left[ \mathcal{M}_{j}\left(\mathbf{k}_{3}\right) \right],\nonumber 
\label{calB1}
\end{eqnarray}
where $\mathrm{Re} \left[ \ldots \right]$ means the real part and $\mathcal{B}_{\zeta}^{\mathrm{local}}\left(\mathbf{k}_{1},\mathbf{k}_{2},\mathbf{k}_{3}\right)$
is defined similarly to Eq.~(\ref{eq:curlyB_equil_def}) \begin{equation}
\mathcal{B}_{\zeta}^{\mathrm{local}}\left(\mathbf{k}_{1},\mathbf{k}_{2},\mathbf{k}_{3}\right)\equiv\frac{k_{1}^{3}k_{3}^{3}}{4\pi^{4}}B_{\zeta}^{\mathrm{local}}\left(\mathbf{k}_{1},\mathbf{k}_{2},\mathbf{k}_{3}\right).\end{equation}
Then, the nonlinearity parameter $f_{\mathrm{NL}}^{\mathrm{local}}$ in the
squeezed configuration becomes\begin{equation}
\frac{6}{5}f_{\mathrm{NL}}^{\mathrm{local}}=\frac{\mathcal{B}_{\zeta}^{\mathrm{local}}\left(\mathbf{k}_{1},\mathbf{k}_{2},\mathbf{k}_{3}\right)}{2\mathcal{P}_{\zeta}^{\mathrm{iso}}\left(k_{1}\right)\mathcal{P}_{\zeta}^{\mathrm{iso}}\left(k_{3}\right)}.
\label{fNLloc}\end{equation}

Having defined our notation, in the following sections we will consider
two particular examples. But before that, we wish to point out an
important subtlety concerning the vector field $A_{\mu}$. By $A_{i}$
we refer to the spatial components of the \emph{physical} vector field
$A_{i}=B_{i}/a$, where $a$ is the scale factor in FRW universe.
The field $B_{i}$, which enters the Lagrangian, is the \emph{comoving}
vector field with the expansion of the universe factored out. 
In FRW spacetime the temporal components of the physical and comoving
fields are the same, $A_{0}=B_{0}$ \citep{Dimopoulos2006,Dimopoulos2007}.

\section{Anisotropy in the vector curvaton model}\label{vecurvsec}

In this section we study the non-Gaussianity in the curvature perturbation
for a model of slow roll
inflation with an additional $U\left(1\right)$ vector field which
decays some time after reheating and contributes to the total curvature
perturbation following the curvaton mechanism \citep{Lyth_Wands(2002)}.
For this contribution to be non-negligible, the vector field must
undergo particle production during inflation and obtain a superhorizon
spectrum of perturbations. But the massless $U\left(1\right)$ field
is conformally invariant and, consequently, its quantum fluctuations 
are not amplified during inflation. This means that in order for such
a vector field to undergo particle production we have to brake its
conformality \citep{TurnerWidrow1988}. Another problem is that a
dominant homogeneous (homogenized by inflation) vector field could
make the expansion of the Universe strongly anisotropic, which is
in contradiction with observations.

One way out of this problem can be the curvaton mechanism 
\citep{Lyth_Wands(2002),Moroi_Takahashi(2001),Linde_Mukhanov(1997)}.
The usual curvaton scenario incorporates two scalar fields: one that
drives inflation and another one, called curvaton, which produces the
curvature perturbation. During inflation the curvaton is subdominant.
However, after reheating the universe is radiation dominated and its
energy density is diluted as $\rho_{r}\propto a^{-4}$. If the energy
density of the curvaton field decreases slower than $a^{-4}$, at
some moment it can dominate (or nearly dominate) the Universe and
impose its own curvature perturbation. This is the basic idea of the
curvaton mechanism.

Here we consider a massive vector field acting as the curvaton. Before
dominating the vector curvaton field is rapidly oscillating in a quasi-harmonic
manner. As shown in Ref.~\citep{Dimopoulos2006}, the oscillating
vector field behaves as a pressureless \emph{isotropic} fluid and
can dominate without generating a large-scale anisotropy. 

\subsection{The generic treatment}

Here we obtain analytic expressions for the non-linearity parameter 
$f_{\rm NL}$ without assuming a specific vector curvaton model.
In contrast to the original curvaton idea we include as well perturbations 
generated during inflation by the light scalar field (for a similar study in 
the scalar curvaton case see 
Refs.~\citep{Lazarides_etal(2004),Ichikawa_etal(2008)}).

When some time after reheating the mass of the vector field becomes
bigger than the Hubble parameter, the field starts to oscillate. In
Ref.~\citep{Dimopoulos_etal(2008)} it was shown that for oscillating
vector field $N_{i}^{A}$ and $N_{ij}^{A}$ are equal to
\begin{equation}
N_{i}^{A}=\frac{2}{3}r\frac{A_{i}}{A^{2}},\label{eq:NAi_def}\end{equation}
\begin{equation}
N_{ij}^{A}=\frac{2}{3}r\frac{\delta_{ij}}{A^{2}},\label{eq:NAij_def}\end{equation}
 where $A\equiv\left|\mathbf{A}\right|$ is evaluated just before
the vector field decays and the parameter $r$ is defined as 
\begin{equation}
r\equiv\frac{3\rho_{A}}{3\rho_{A}+4\rho_{r}}=\frac{3\Omega_{A}}{4-\Omega_{A}}
\end{equation}
 with $\rho_{A}$ being the energy density of the vector field just
before its decay (taken to be sudden%
), 
$\Omega_{A}\equiv\rho_{A}/\rho$
and $\rho=\rho_{A}+\rho_{r}$. 
Using Eq.~(\ref{eq:NAi_def}) the isotropic part of the total power
spectrum in Eq.(\ref{eq:Piso_definition}) becomes 
\begin{equation}
\mathcal{P}_{\zeta}^{\mathrm{iso}}
=N_{\phi}^{2}\mathcal{P}_{\phi}
\left(1+\beta\,\frac{\mathcal{P}_{+}
}{\mathcal{P}_{\phi}
}\right),\label{eq:P_iso_delta}
\end{equation}
 where we defined
\begin{equation}
\beta\equiv\left(\frac{N_{A}}{N_{\phi}}\right)^2.
\end{equation}
 Then the vector part of the bispectrum for equilateral configuration
in Eq.(\ref{eq:BA_def}) reduces to \begin{eqnarray}
\mathcal{B}_{A}^{\mathrm{equil}}\left(\mathbf{k}_{1},\mathbf{k}_{2},\mathbf{k}_{3}\right) & = & \left(\frac{2}{3}\frac{r}{A}\right)^{3}\frac{1}{A}\mathcal{P}_{+}\left(k_{1}\right)\mathcal{P}_{+}\left(k_{2}\right)\biggl\{1+p\left(k_{1}\right)A_{1}^{2}+p\left(k_{2}\right)A_{2}^{2}+A_{1}A_{2}\left[q\left(k_{1}\right)q\left(k_{2}\right)-\frac{1}{2}p\left(k_{1}\right)p\left(k_{2}\right)\right]+\nonumber \\
 &  & \left.+i\sqrt{\frac{3}{4}-\left(A_{1}^{2}+A_{1}A_{2}+A_{2}^{2}\right)}\left[A_{1}p\left(k_{1}\right)q\left(k_{2}\right)-A_{2}p\left(k_{2}\right)q\left(k_{1}\right)\right]+\frac{1}{2}q\left(k_{1}\right)q\left(k_{2}\right)\right\} +\mathrm{c.p.}\label{eq:M^2_equilateral}\end{eqnarray}
In the above we used the notation $A_{1}\equiv\hat{\mathbf{A}}\cdot\hat{\mathbf{k}}_{1}$
etc., where $\hat{\mathbf{A}}=\mathbf{A}/A$. Because the configuration
of wavevectors $\hat{\mathbf{k}}_{1}$, $\hat{\mathbf{k}}_{2}$ and
$\hat{\mathbf{k}}_{3}$ is equilateral, with the angle between any
two of them being $2\pi/3$, we find $\hat{\mathbf{k}}_{1}\cdot\hat{\mathbf{k}}_{2}=\hat{\mathbf{k}}_{1}\cdot\hat{\mathbf{k}}_{3}=\hat{\mathbf{k}}_{2}\cdot\hat{\mathbf{k}}_{3}=-\frac{1}{2}$.
Eq.(\ref{eq:M^2_equilateral}) simplifies further if we consider a
scale invariant power spectrum and the expression for $f_{\mathrm{NL}}^{\mathrm{equil}}$
becomes:\begin{equation}
\frac{6}{5}f_{\mathrm{NL}}^{\mathrm{equil}}=\beta^{2}\mathcal{P}_{+}^{2}\,\frac{3}{2r}\,\frac{\left(1+\frac{1}{2}q^{2}\right)+\left[p+\frac{1}{8}\left(p^{2}-2q^{2}\right)\right]A_{\bot}^{2}}{\left(\mathcal{P}_{\phi}+\beta\,\mathcal{P}_{+}\right)^{2}},\label{eq:M^2_vCurv_scaleInv}
\end{equation}
where we have taken into account that the non-Gaussianity generated
during the single field inflation is negligible. The quantity $A_{\bot}\leq1$
is the modulus of the projection of the unit vector $\hat{\mathbf{A}}$
onto the plane containing the three vectors $\hat{\mathbf{k}}_{1}$, $\hat{\mathbf{k}}_{2}$
and $\hat{\mathbf{k}}_{3}$. The calculation of $A_{\bot}$ in the equilateral
configuration is explained in more detail in the Appendix.

For the squeezed configuration the bispectrum from the vector field
perturbation in Eqs.~(\ref{eq:Bs_local}) becomes
\begin{eqnarray}
\mathcal{B}_{A}^{\mathrm{local}}\left(\mathbf{k}_{1},\mathbf{k}_{2},\mathbf{k}_{3}\right) & = & 
2\left(\frac{2}{3}\frac{r}{A}\right)^{3}\frac{1}{A}\mathcal{P}_{+}\left(k_{1}\right)\mathcal{P}_{+}\left(k_{3}\right)
\left[1 + p\left(k_{1}\right)A_{1}^{2} + p\left(k_{3}\right)A_{3}^{2}\right] 
.\end{eqnarray}
Working as in the equilateral case, we find that the
non-linearity parameter for the scale invariant power spectra is
\begin{equation}
\frac{6}{5}f_{\mathrm{NL}}^{\mathrm{local}}=\beta^{2}\mathcal{P}_{+}^{2}\,\frac{3}{2r}\,\frac{1+ p A_{\bot}^{2}}{\left(\mathcal{P}_{\phi}+\beta\,\mathcal{P}_{+}\right)^{2}}.\label{eq:fNL_scaleInv_vCurvaton_local}
\end{equation}
In this equation $\varphi$ is the angle between the vectors $\mathbf{k}_{1}$
and $\mathbf{A}_{\bot}$.

As one can see from the above equations, $f_{\mathrm{NL}}$ is, in
general, dependent on $A_{\bot}$, in both configurations. This means
that $f_{\mathrm{NL}}$ is anisotropic and that the amount of non-Gaussianity
is correlated with the statistical anisotropy. However, from 
Eqs.~(\ref{eq:Ppm_def}), (\ref{eq:g_h_def}) and (\ref{eq:miu_def}) 
it is clear that, if particle
production is isotropic (i.e. \mbox{$\mathcal{P}_{0}=\mathcal{P}_{+}$}
and \mbox{$\mathcal{P}_{-}=0$}) then \mbox{$p=q=0$} and the
above expressions for $f_{\mathrm{NL}}^{\mathrm{equil}}$ and $f_{\mathrm{NL}}^{\mathrm{local}}$
become isotropic too and both reduce to 
\mbox{$f_{\mathrm{NL}}
=5/4r$}
as in the scalar curvaton scenario, where we have assumed that \mbox{$\mathcal{P}_{\phi}\ll\mathcal{P}_{+}$},
i.e. that the dominant contribution to the curvature perturbation
is due to the vector curvaton field only.

\subsection{$\boldsymbol{f_{\mathrm{NL}}}$ for non-minimally coupled vector
curvaton\label{sub:fNL-non-minimal-vCurv}}

In Ref.~\citep{Dimopoulos2006} it was shown that a vector field
can attain a scale invariant perturbation spectrum if it's mass during
inflation is equal to $m^{2}=-2H^{2}$. One way to achieve a negative
mass squared of this magnitude 
is to introduce a non-minimal coupling of the vector
field to gravity of the form $\frac{1}{6}RB^{\mu}B_{\mu}$, where $R$ is the 
Ricci scalar. The idea
of such a non-minimally coupled vector curvaton was introduced in 
Ref.~\citep{Dimopoulos_Karciauskas(2008)}.
In that paper and in Ref.~\citep{Dimopoulos_etal(2008)} it was shown
that the power spectra for different polarizations are\begin{equation}
\mathcal{P}_{+}=\left(\frac{H}{2\pi}\right)^{2},\quad\mathcal{P}_{-}=0\quad\mathrm{and}\quad\mathcal{P}_{0}=2\left(\frac{H}{2\pi}\right)^{2}.\label{eq:Ps_RA^2}\end{equation}

One notices that the parity conserving transverse power spectrum and the power
spectrum generated during the single scalar field inflation are equal,
i.e. $\mathcal{P}_{+}=\mathcal{P}_{\phi}$. Thus the isotropic part
of the curvature perturbation spectrum can be written as
\begin{equation}
\mathcal{P}_{\zeta}^{\mathrm{iso}}
=\mathcal{P}_{\phi}N_{\phi}^{2}\left(1+\beta\right).
\end{equation}
 While the anisotropy parameter from Eq.(\ref{eq:miu_def}) becomes
\begin{equation}
g=\frac{\beta}{1+\beta}.\label{eq:miu_vCurvaton}
\end{equation}
 Using Eq.(\ref{eq:Ps_RA^2}) we find
\begin{equation}
p=1\quad\mathrm{and}\quad q=0.\label{eq:g_h_RA^2}\end{equation}
Thus, the anisotropy in the vector field is rather strong, which means that it 
will have to remain subdominant, i.e. \mbox{$\Omega_A\ll 1$}.
Using this and 
Eq.~(\ref{eq:M^2_vCurv_scaleInv}), the $f_{\mathrm{NL}}^{\mathrm{equil}}$
for the non-minimally coupled vector curvaton is found to be
\begin{eqnarray}
\frac{6}{5}f_{\mathrm{NL}}^{\mathrm{equil}}=\frac{\beta^{2}}{4\Omega_{A}}\left(8+9A_{\bot}^{2}\right).\label{eq:fNL_vCurvaton}\end{eqnarray}
Similarly,
$f_{\mathrm{NL}}^{\mathrm{local}}$ for the squeezed configuration
in Eq.~(\ref{eq:fNL_scaleInv_vCurvaton_local}) is\begin{equation}
\frac{6}{5}f_{\mathrm{NL}}^{\mathrm{local}}=2\frac{\beta^{2}}{\Omega_{A}}\left(1+A_{\bot}^{2}\right)\label{eq:fNL_vCurvaton_local}\end{equation}

Since 
\mbox{$\calp_+=\frac12\calp_0=\calp_\phi=\left(\frac{H}{2\pi}\right)^2$}, 
for the typical values of the perturbations we have
\mbox{$\delta\phi\sim\delta A_i\sim H$}. This means that, in order for the 
vector field contribution to be subdominant, we require \mbox{$N_A\ll N_\phi$}
(c.f. Eq.~(\ref{eq:zeta_x})). Hence, \mbox{$\beta\ll 1$} and 
\mbox{$g\simeq\beta$}. Thus, in view of Eqs.~(\ref{eq:fNL_vCurvaton}) 
and (\ref{eq:fNL_vCurvaton_local}), we see that 
\mbox{$f_{\mathrm{NL}}\sim g^2/\Omega_A$}. Therefore, we find that the 
non-Gaussianity is determined by the magnitude of the statistical anisotropy.

This prediction is valid in the regime $|\delta A/A|\ll1$ which corresponds
to $\Omega_{A}^{2}\gsim\calpz\beta$, which implies 
\mbox{$f_{\mathrm{NL}}\lsim g^{3/2}/\sqrt{\calp_\zeta}$}. 
For smaller $\Omega_{A}$,
the contribution of the vector field perturbation to $\zeta$ is of
order $\Omega_{A}[\delta A/(\overline{\delta A^{2}})\half]$. In other
words, it is of order $\Omega_{A}$ and is the square of a Gaussian
quantity. The resulting prediction for its contribution to $\fnl$
would be given by a 1-loop formula which has not been evaluated at
the time of writing.

\section{Anisotropy generated at the end of inflation\label{sec:fNL-end-of-inflation}}

As another example, we consider the generation of an anisotropic power
spectrum at the end of inflation. The idea is based on Ref.~\citep{Lyth(2005a)}
where it was shown that in hybrid inflation models the generation
of curvature perturbations can be realized due to inhomogeneous end
of inflation. Yokoyama and Soda~\citep{Yokoyama_Soda(2008)} used
this idea to generate the anisotropic contribution to the total curvature
perturbations. In their model the anisotropy is generated at the end
of inflation due to the vector field coupling with the waterfall field.
In this section we calculate the non-Gaussianity of the model in Ref.~\citep{Yokoyama_Soda(2008)}
using our notation in section \ref{sec:general_fomulae}.

In this model there are two components of the curvature perturbation:
one generated during inflation and an anisotropic one, generated by
a vector field at the end of inflation:\begin{equation}
\zeta=\zeta_{\mathrm{inf}}+\zeta_{\mathrm{end}}.\end{equation}
 The first component is due to the perturbation generated during inflation.
The second component is due to the perturbation of the vector field.
Ref.~\citep{Yokoyama_Soda(2008)} considers a massless, $U\left(1\right)$
vector field. Without parity violating terms the power spectra for
left handed and right handed polarizations are equal, while the longitudinal
polarization is absent for a massless field. In this situation we
find that parameters $p\left(k\right)$ and $q\left(k\right)$ from
Eq.(\ref{eq:g_h_def}) become\begin{equation}
p=-1\quad\mathrm{and}\quad q=0.\label{eq:g_h_at_end_infl}\end{equation}

The conformal invariance of the $U\left(1\right)$ vector field is
broken through a non-canonical kinetic function of the form $f^{2}\left(t\right)F_{\mu\nu}F^{\mu\nu}$,
where $F_{\mu\nu}=\partial_{\mu}B_{\nu}-\partial_{\nu}B$ is the field
strength tensor and $B_{\mu}$ - comoving vector field. This form
of conformal invariance braking was considered in many papers (e.g.
Refs.~\citep{Dimopoulos2007,Bamba_Yokoyama(2004),BambaSasaki2007,Martin_Yokoyama(2008),Seery(2008)})
where it was found that a scale invariant perturbation spectrum is
obtained if $f\propto a$:\begin{equation}
\mathcal{P}_{+}=\left(\frac{H}{2\pi f}\right)^{2}=\mathcal{P}_{\phi}f^{-2}.\end{equation}
 So the isotropic part in Eq.(\ref{eq:Piso_definition}) of the power
spectrum becomes
\begin{equation}
\mathcal{P}_{\zeta}^{\mathrm{iso}}=\mathcal{P}_{\phi}N_{\phi}^{2}\left(1+\beta\right).\label{eq:Piso_YS}
\end{equation}
 This is of the same form as with the vector curvaton model but with
different $\beta$: \begin{equation}
\beta=\left(\frac{N_{A}}{N_{\phi}f}\right)^{2}.\end{equation}
 The anisotropy parameter in Eq.(\ref{eq:miu_def}) in this model
becomes\begin{equation}
g=-\frac{\beta}{1+\beta}.\end{equation}
 Taking into account Eq.(\ref{eq:g_h_at_end_infl}), the vector $\mathcal{M}_{i}\left(\mathbf{k}\right)$
in Eq.(\ref{eq:M_def}) reduces to the simple form
\begin{equation}
\mbox{\boldmath $\mathcal{M}$}\left(\mathbf{k}\right)=
N_{A}\mathcal{P}_{\phi}f^{-2}\left[\hat{\bf N}^{A}-\hat{\bf k}\left(\hat{\mathbf{N}}^{A}\cdot\hat{\mathbf{k}}\right)\right].
\label{eq:Mi_YS_1}
\end{equation}
 To calculate $f_{\mathrm{NL}}$ we 
consider a specific example
of hybrid inflation with the potential \begin{equation}
V\left(\phi,\chi,B^{\mu}\right)=V_{0}+\frac{1}{2}m_{\phi}^{2}\phi^{2}-\frac{1}{2}m_{\chi}^{2}\chi^{2}+\frac{1}{4}\lambda\chi^{4}+\frac{1}{2}\lambda_{\phi}\phi^{2}\chi^{2}+\frac{1}{2}\lambda_{A}\chi^{2}B^{\mu}B_{\mu},\end{equation}
 where $\phi$ is the inflaton and $\chi$ is the waterfall field.
The effective mass of the waterfall field
for this potential is \begin{eqnarray}
m_{\mathrm{eff}}^{2} & = & -m_{\chi}^{2}+\lambda_{\phi}\phi^{2}-\lambda_{A}A_{i}A_{i},\label{eq:m^2eff_def}\end{eqnarray}
 where Einstein summation is assumed and we used $A_{i}\equiv B_{i}/a$
and the Coulomb gauge in which $A_{0}=0$ and $\partial_{i}A^{i}=0$.
Inflation ends when the inflaton reaches a critical value $\phi_{c}$
where the effective mass of the waterfall field becomes tachyonic.
But one can see from Eq.(\ref{eq:m^2eff_def}) that the critical value
is a function of the vector field $\phi_{c}=\phi_{c}\left(A\right)$.
With this in mind the vectors $N_{i}^{A}$ and $N_{ij}^{A}$ can be
readily calculated:\begin{equation}
N_{i}^{A}=\frac{\partial N}{\partial\phi_{c}}\frac{\partial\phi_{c}}{\partial A_{i}}=N_{c}\frac{\lambda_{A}}{\lambda_{\phi}}\frac{A_{i}}{\phi_{c}},\end{equation}
 and\begin{equation}
N_{ij}^{A}=\frac{\partial N}{\partial\phi_{c}}\frac{\partial^{2}\phi_{c}}{\partial A^{i}\partial A^{j}}+\frac{\partial^{2}N}{\partial\phi_{c}^{2}}\frac{\partial\phi_{c}}{\partial A_{i}}\frac{\partial\phi_{c}}{\partial A_{j}}=\frac{N_{A}^{2}}{\phi_{c}N_{c}}\left(C^{2}\delta_{ij}-\hat{A}_{i}\hat{A}_{j}\right),\end{equation}
 where we have defined \begin{equation}
N_{c}=\frac{\partial N}{\partial\phi_{c}}\quad\mathrm{and}\quad C\equiv\sqrt{\frac{\lambda_{\phi}}{\lambda_{A}}}\frac{\phi_{c}}{A},\end{equation}
where $A$ is evaluated at the end of inflation
 and we used the fact that $N_{cc}/N_{c}^{2}\sim N_{\phi\phi}/N_{\phi}^{2}\sim\mathcal{O}\left(\epsilon\right)$
under the slow roll approximation \citep{Lyth(2005a)}, where $\epsilon$
is the slow roll parameter defined as 
$\epsilon\equiv\frac{1}{2}M_{\mathrm{P}}^{2}\left(V'/V\right)^{2}$,
with the prime denoting derivatives with respect to the inflaton.
As mentioned earlier the total of perturbations consists
of two components: perturbations of the scalar and vector fields.
This gives the following bispectrum in the equilateral configuration
\begin{eqnarray}
\mathcal{B}_{\zeta}^{\mathrm{equil}}\left(\mathbf{k}_{1},\mathbf{k}_{2},\mathbf{k}_{3}\right) & = & \mathcal{B}_{\phi}^{\mathrm{equil}}\left(\mathbf{k}_{1},\mathbf{k}_{2},\mathbf{k}_{3}\right)+\mathcal{B}_{A}^{\mathrm{equil}}\left(\mathbf{k}_{1},\mathbf{k}_{2},\mathbf{k}_{3}\right)=\nonumber \\
& = & 3\mathcal{P}_{\phi}^{2}N_{\phi}^{2}N_{\phi\phi}+\left[\mathcal{M}_{i}\left(\mathbf{k}_{1}\right)N_{ij}^{A}\mathcal{M}_{j}\left(\mathbf{k}_{2}\right)+\mathrm{c.p.}\right]=\nonumber \\
 & = & \mathcal{P}_{\phi}^{2}N_{\phi}^{4}\frac{\delta^{2}}{N_{c}\phi_{c}}3\left[\left(C^{2}-1\right)-\left(\frac{7}{8}C^{2}-1\right)A_{\bot}^{2}-\frac{3}{16}A_{\bot}^{4}\right].\label{eq:B_end_final}\end{eqnarray}
 The mixed term $\mathcal{B}_{\phi A}^{\mathrm{equil}}$ is absent
from Eq.(\ref{eq:B_end_final}) because in this model $N_{\phi i}^{A}=0$.
By inserting Eq.(\ref{eq:Piso_YS})
into (\ref{eq:B_end_final}) we obtain \begin{equation}
\frac{6}{5}f_{\mathrm{NL}}^{\mathrm{equil}}=\eta g^{2}\left[\left(C^{2}-1\right)-\left(\frac{7}{8}C^{2}-1\right)A_{\bot}^{2}-\frac{3}{16}A_{\bot}^{4}\right],\label{eq:fNL-end-of-infl-equil}\end{equation}
 where the slow parameter $\eta$ is equal to $\eta=m_{\phi}^{2}M_{\mathrm{P}}^{2}/V_{0}$
and $M_{\mathrm{P}}N_{c}=1/\sqrt{2\epsilon_{c}}$, with $\epsilon_{c}$
being the slow roll parameter at the end of inflation. 
Similarly, for the squeezed configuration we find\begin{equation}
\frac{6}{5}f_{\mathrm{NL}}^{\mathrm{local}}=\eta g^{2}\left[\left(C^{2}-1\right)\left(1-A_{\bot}^{2}\right)-\frac{1}{4}(\sin\varphi)^2A_{\bot}^{4}\right].\label{eq:fNL-end-of-infl-local}\end{equation}

Again, we find that $f_{\mathrm{NL}}^{\mathrm{equil}}$ and $f_{\mathrm{NL}}^{\mathrm{local}}$
are functions of $A_{\bot}$, i.e. they are anisotropic and correlated
with the statistical anisotropy. Also the level of non-Gaussianity
is proportional to the anisotropy parameter squared, 
$f_{\mathrm{NL}}
\propto g^{2}$,
as in the non-minimally coupled vector curvaton model.

\section{Discussion}

In this paper we have calculated the amount of the non-Gaussianity
generated by the anisotropic part of the curvature perturbation. We
have also considered two specific models to generate the anisotropic
curvature perturbation from the vector fields and calculated the 
non-Gaussianity
in detail for those models. The results were given for the equilateral
and squeezed configurations in Eqs.~(\ref{eq:fNL_vCurvaton}) and
(\ref{eq:fNL_vCurvaton_local}) for the model of non-minimally coupled
vector curvaton and in Eqs.~(\ref{eq:fNL-end-of-infl-equil}), (\ref{eq:fNL-end-of-infl-local})
for the end-of-inflation scenario. We have shown that $f_{\mathrm{NL}}$
generated by the vector field is anisotropic and that it is
correlated with the amount and direction of the statistical
anisotropy. 

Although in some specific models it would be possible to generate
the statistically isotropic curvature perturbation only from the vector
field, however, in general, perturbations generated by the vector
field are anisotropic. From observations the bound on statistically
anisotropic contribution to the total power spectrum is constrained
to be less than about $10\,\%$. In this case one can estimate the maximal
$f_{\mathrm{NL}}$ caused by the statistically anisotropic curvature
perturbation on fairly general grounds. 

Let us assume that the non-Gaussianity is produced solely due to the vector 
field perturbations. If this is so then Eqs.~(\ref{eq:fNL_definition1}) and 
(\ref{fNLloc}) suggest \mbox{$f_{\rm NL}\sim{\cal B}_\zeta/\calp_\zeta^2$},
where we consider that the anisotropic contribution to the curvature 
perturbation is subdominant, i.e. 
\mbox{$\calp_\zeta\simeq\calp_\zeta^{\rm iso}$}.
According to Eqs.~(\ref{calB}) and (\ref{calB1}) we have 
\mbox{${\cal B}_\zeta\sim{\cal M}^2N_{AA}$}, where 
\mbox{${\cal M}\equiv |\mbox{\boldmath ${\cal M}$}|$} and 
\mbox{$N_{AA}\equiv ||N^A_{ij}||$}. Thus, we have
\mbox{$f_{\rm NL}\sim{\cal M}^2N_{AA}/\calp_\zeta^2$}.

Now, ${\cal M}$ depends on the mechanism which breaks the conformality of the
vector field and is responsible for the generation of its superhorizon 
perturbation spectrum. If this mechanism does not introduce additional mass
scales, then, on dimensional grounds, we expect the anisotropy in the vector 
field perturbation to be of order 
unity, i.e. \mbox{$|p|,|q|\sim{\cal O}(1)$}, barring cancellations such as
due to parity invariance (which results in $q=0$) or an isotropic particle 
production, which gives $p=q=0$ and generates no statistical anisotropy. 
In our examples in Secs.~\ref{vecurvsec} and \ref{sec:fNL-end-of-inflation}
we indeed analysed such a situation, where the effective mass-squared of the
vector field during inflation \mbox{$m^2=\frac16 R=-2H^2$} or the time 
dependence of the kinetic function \mbox{$\dot f/f=H$} were both determined
by the dynamics of the expansion and given by $H$, the only scale in the 
theory. If the anisotropy in the vector 
field perturbation is of order unity, then Eq.~(\ref{eq:M_def}) gives
\mbox{${\cal M}\sim\calp_A N_A$}, where \mbox{$\calp_A=2\calp_++\calp_0$} is
the power spectrum of the total vector field perturbation given by
\mbox{$\calp_A=\frac{k^3}{2\pi^2}\sum_i|\delta A_i|^2$} 
in the superhorizon limit. Putting the above together we obtain
\be
f_{\rm NL}\sim\frac{{\cal B}_\zeta}{\calp_\zeta^2}
\sim\frac{{\cal M}^2N_{AA}}{\calp_\zeta^2}
\sim\frac{\calp_A^2 N_A^2N_{AA}}{\calp_\zeta^2}.
\ee

Since we are working in the regime where $\left|\delta A/A\right|\ll1$ we
expect the higher order contribution to $\zeta$ in Eq.~(\ref{eq:zeta_x})
from the vector field to be subdominant, i.e. 
\mbox{$N_A\delta A>N_{AA}\delta A^2$}. Considering that the typical value of
the vector field perturbation is \mbox{$\delta A\sim\sqrt{\calp_A}$}, we
obtain the bound \mbox{$f_{\rm NL}<(\calp_A^{1/2}N_A)^3/\calp_\zeta^2$}. As 
evident from Eq.~(\ref{eq:zeta_x}), the contribution of the vector field to
$\zeta$ is given by 
\mbox{$\zeta_A\sim\sqrt{\calp_{\zeta_A}}\sim N_A\sqrt{\calp_A}$}, 
where $\mathcal{P}_{\zeta_{A}}$ is the power spectrum of the anisotropic
curvature perturbation. With this in 
mind, the upper bound to $f_{\mathrm{NL}}$ becomes
\begin{equation}
f_{\mathrm{NL}}\,\mathcal{P}_{\zeta}^{1/2}<
\left(\frac{\mathcal{P}_{\zeta_{A}}}{\mathcal{P}_{\zeta}}\right)^{3/2}.
\end{equation}
Because
the vector field contribution to the total curvature perturbation
must be subdominant, Eq.~(\ref{eq:Pzeta_definition}) suggests that
the anisotropy of the curvature perturbation is
\mbox{$g\sim\mathcal{P}_{\zeta_{A}}/\mathcal{P}_{\zeta}$}.
Using this and also that $\mathcal{P}_{\zeta}\approx5\times 10^{-5}$, 
we find that the maximum value of $f_{\mathrm{NL}}$
generated by the statistically anisotropic contribution to the curvature
perturbation has to be
\begin{equation}
f_{\mathrm{NL}}^{\mathrm{max}}\sim10^{3}\left(\frac{g}{0.1}\right)^{3/2}.
\end{equation}
Our results in Eqs.~(\ref{eq:fNL_vCurvaton}), (\ref{eq:fNL_vCurvaton_local}),
(\ref{eq:fNL-end-of-infl-equil}) and (\ref{eq:fNL-end-of-infl-local})
apply if $f_{\mathrm{NL}}$ is below this value.

\section{Conclusions}

It is clear that the study of vector field contributions to the primordial
curvature perturbation is just beginning. Given a scale invariant
and Gaussian vector field perturbation, the calculation of $\zeta$
from the $\delta N$ formalism is straightforward and should now be
done for the full range of scenarios that have already been explored
for the contribution of scalar field perturbations. Also, the Feynman
graph formalism available for the scalar field case should be generalized
to cover the vector field contributions. At a deeper level, one also
wishes to understand how the scale invariant perturbation can be generated.
One would also like a fuller understanding of the generation of perturbations
from the vacuum fluctuation when the expansion of the unperturbed
Universe is anisotropic, since that can easily happen in the presence
of vector fields.

For the moment though, it is most urgent to confront specific
predictions for the form of the anisotropy with observation. Specifically,
we want to know what constraint is placed by observation on a contribution
to $\fnl$ of the form in 
Eqs.~(\ref{eq:fNL_vCurvaton}), (\ref{eq:fNL_vCurvaton_local})
or Eqs.~(\ref{eq:fNL-end-of-infl-equil}), (\ref{eq:fNL-end-of-infl-local}). 
It seems 
quite possible that with the latter prediction,
valid for the squeezed configuration, one might find a nonzero value
at better than the 2-$\sigma$ level that is already found for the
isotropic case. Such a finding, if confirmed, would be a smoking gun
for a vector field contribution to the curvature~perturbation.

\acknowledgements
This work was supported (in part) by the European Union through the 
Marie Curie Research and Training Network "UniverseNet" 
(MRTN-CT-2006-035863) and by STFC (PPARC) Grant PP/D000394/1.

\newpage{}

\appendix

\section*{Appendix}

\section*{Calculation of $\boldsymbol{A_{\bot}}$ in the equilateral configuration}

First note that in the equilateral configuration $\hat{\mathbf{k}}_{1}+\hat{\mathbf{k}}_{2}=-\hat{\mathbf{k}}_{3}$.
This gives $A_{1}+A_{2}=-A_{3}$ and\begin{equation}
\begin{array}{l}
A_{1}^{2}+A_{2}^{2}+A_{3}^{2}=2\left(A_{1}^{2}+A_{1}A_{2}+A_{2}^{2}\right);\\
A_{1}A_{2}+A_{2}A_{3}+A_{3}A_{1}=-\left(A_{1}^{2}+A_{1}A_{2}+A_{2}^{2}\right);\\
A_{1}^{2}A_{2}^{2}+A_{2}^{2}A_{3}^{2}+A_{3}^{2}A_{1}^{2}=\left(A_{1}^{2}+A_{1}A_{2}+A_{2}^{2}\right)^{2}.\end{array}\label{eq:As}
\end{equation}
 Let us define a vector $\mathbf{A}_{\bot}$ which is the projection
of $\hat{\mathbf{A}}$ to the plane containing vectors $\hat{\mathbf{k}}_{1}$,
$\hat{\mathbf{k}}_{2}$ and $\hat{\mathbf{k}}_{3}$. Then the scalar
product of these vectors and $\hat{\mathbf{A}}$ is the same as the
product with $\mathbf{A}_{\bot}$: 
\begin{equation}
\hat{\mathbf{A}}\cdot\hat{\mathbf{k}}_{a}=\mathbf{A}_{\bot}\cdot\hat{\mathbf{k}}_{a},
\end{equation}
where $a=1,2,3$.

Without loss of generality we can assume that the angle between $\mathbf{A}_{\bot}$
and $\hat{\mathbf{k}}_{1}$ is $\varphi$:
\begin{equation}
A_{1}\equiv\hat{\mathbf{A}}\cdot\hat{\mathbf{k}}_{1}=\mathbf{A}_{\bot}\cdot\hat{\mathbf{k}}_{1}=A_{\bot}\cos\varphi,\label{eq:A1}
\end{equation}
 where $A_{\bot}=\left|\mathbf{A}_{\bot}\right|$. In equilateral
configuration the angle between vectors $\hat{\mathbf{k}}_{1}$ and
$\hat{\mathbf{k}}_{2}$ is $2\pi/3$, and $A_{2}$ becomes\begin{equation}
A_{2}\equiv\mathbf{A}_{\bot}\cdot\hat{\mathbf{k}}_{2}=A_{\bot}\cos\left(\varphi+\frac{2\pi}{3}\right)=-A_{\bot}\left(\frac{1}{2}\cos\varphi+\frac{\sqrt{3}}{2}\sin\varphi\right).\label{eq:A2}\end{equation}
 From the last two equations we get\begin{equation}
A_{1}^{2}+A_{1}A_{2}+A_{2}^{2}=\frac{3}{4}A_{\bot}^{2}.\label{eq:Apr_value}\end{equation}
 Putting this result back into Eq.(\ref{eq:As}) we find\begin{equation}
\begin{array}{l}
A_{1}^{2}+A_{2}^{2}+A_{3}^{2}=\frac{3}{2}A_{\bot}^{2};\\
A_{1}A_{2}+A_{2}A_{3}+A_{3}A_{1}=-\frac{3}{4}A_{\bot}^{2};\\
A_{1}^{2}A_{2}^{2}+A_{2}^{2}A_{3}^{2}+A_{3}^{2}A_{1}^{2}=\frac{9}{16}A_{\bot}^{4}.\end{array}\label{eq:A_values}\end{equation}

\end{document}